# Acute lymphoblastic leukemia may develop as a result of rapid transformation of a lymphoblast triggered by repeated bone-remodeling during bone-growth

Jicun Wang-Michelitsch[1]*, Thomas M Michelitsch[2]

[1] Independent researcher

[2] Sorbonne Université, Institut Jean le Rond d'Alembert, CNRS UMR 7190 Paris, France

## Abstract

Acute lymphoblastic leukemia (ALL) and chronic lymphocytic leukemia (CLL) are two major forms of leukemia that arise from lymphoid cells (LCs). ALL occurs mostly in children and CLL occurs mainly in old people. However, the Philadelphia-chromosome-positive ALL ($Ph^+$-ALL) and the Ph-like ALL occur in both children and adults. To understand childhood leukemia/lymphoma, we have recently proposed two hypotheses on the causes and the mechanism of cell transformation of a LC. Hypothesis **A** is**:** repeated bone-remodeling during bone-growth and bone-repair may be a source of cell injuries of marrow cells including hematopoietic stem cells (HSCs), myeloid cells, and LCs. Hypothesis **B** is: a LC may have three pathways on transformation: a slow, a rapid, and an accelerated. We discuss in the present paper the developing mechanisms of ALL and CLL by these hypotheses. Having a peak incidence in young children, ALL may develop mainly as a result of rapid cell transformation of a lymphoblast (or pro-lymphocyte). Differently, $Ph^+$-ALL and Ph-like ALL may develop as results of transformation of a lymphoblast via accelerated pathway. Occurring mainly in adults, CLL may be a result of transformation of a memory B-cell via slow pathway. By causing cell injuries of HSCs and LCs, repeated bone-remodeling during bone-growth and bone-repair may be related to the cell transformation of a LC. In conclusion, ALL may develop as a result of cell transformation of a lymphoblast via rapid or accelerated pathway; and repeated bone-remodeling during bone-growth may be a trigger for the cell transformation of a lymphoblast in a child.

**This paper has the following structure:**

I.   **Introduction**

II.  **Age-specific incidences of acute lymphoblastic leukemia (ALL) and chronic lymphocytic leukemia (CLL)**

   2.1  ALL has a peak incidence in young children
   2.2  $Ph^+$-ALL and Ph-like ALL both occur in older children and adults
   2.3  CLL occurs mainly in old people

III. **Three potential sources of cell injuries of lymphoid cells (LCs)**

   3.1  Repeated bone-remodeling during bone-growth and bone-repair in marrow cavity
   3.2  Long-term thymic involution in thymus
   3.3  Pathogen-infections in lymph nodes (LNs) and lymphoid tissues (LTs)

IV.  **DNA changes are generated and accumulate in cells as a result of repeated cell injuries and repeated cell proliferation**

V.   **A LC may have three pathways on cell transformation: a slow, a rapid, and an accelerated**

VI.  **The age of occurrence of leukemia is determined by the transforming pathway of a LC**

   6.1  A leukemia occurring mainly in adults: via slow pathway
   6.2  A leukemia occurring at any age without increasing incidence with age: via rapid pathway
   6.3  A leukemia occurring at any age with increasing incidence with age: via accelerated pathway

VII. **Development of pediatric ALL may be related to the repeated bone-remodeling during bone-growth**

   7.1  Childhood ALL : as a result of rapid cell transformation of a lymphoblast/pro-lymphocyte
   7.2  $Ph^+$-ALL and Ph-like ALL: as consequences of cell injuries of hematopoietic stem cell (HSCs) and LCs in marrow
   7.3  CLL : as a consequence of cell injuries of LCs in LNs/LTs and that of HSCs/LCs in marrow

VIII. **Conclusions**

## I. Introduction

Acute lymphoblastic leukemia (ALL) and chronic lymphocytic leukemia (CLL) are two major forms of lymphoid leukemia originated from lymphoid cells (LCs). "LCs" include all the cells in lymphoid linage at different developing stages. In updated WHO classification of lymphoid neoplasms and acute leukemia in 2016, some subtypes of ALL are renamed by their specific DNA changes. For example, some ALLs are named as B-ALL with t(9;22) (called also BCR-ABL1 or Philadelphia-chromosome-positive ALL ($Ph^+$-ALL)), BCR-ABL1-like ALL (called also Ph-like ALL), or B-ALL with hyperdiploidy (Jaffe, 2017). This novel classification of ALL is important in clinic for making a more precise diagnosis and prognosis for an ALL. 85% of ALLs are B-cell ALL, thus ALLs can be classified in three groups: $Ph^+$-ALL, Ph-like ALL, and B-other ALL. The ALL cases that are neither Ph-positive nor Ph-like belong to the group of B-other ALL (Roberts, 2014).

Adult CLL is thought to be a consequence of long-term repeated exposures of LCs to radiation and/or toxic chemicals in environment. However, the causing factor for childhood ALL is so far unknown. To understand pediatric leukemia and lymphoma, we have recently proposed two hypotheses: **A.** repeated bone-remodeling during bone-growth and bone-repair may be a source of injuries of blood forming cells in marrow; and **B.** a LC may have three pathways on transformation: a slow, a rapid, and an accelerated (Wang-Michelitsch, 2018a and 2018b). In the present paper, we will discuss the developing mechanisms of ALL and CLL by these hypotheses. We aim to show by our discussion that, ALL may develop as a result of cell transformation of a lymphoblast via rapid or accelerated pathway; and repeated bone-remodeling during bone-growth may be a trigger for the cell transformation of a lymphoblast in a child.

We use the following abbreviations in this paper:

| | |
|---|---|
| AML: acute myeloid leukemia | LL: lymphoid leukemia |
| ALL: acute lymphoblastic leukemia | LN: lymph node |
| CC: chromosomal change | LT: lymphoid tissue |
| CML: chronic myeloid leukemia | MECC: mild-effect chromosomal change |
| CLL: chronic lymphocytic leukemia | NCC: numerical chromosome change |
| EBV: Epstein-Barr virus | PDM: point DNA mutation |
| GECC: great-effect chromosomal change | $Ph^+$-ALL: Philadelphia-chromosome-positive ALL |
| HSC: hematopoietic stem cell | SCC: structural chromosome change |
| IECC: Intermediate-effect chromosomal change | SLL: small lymphocyte lymphoma |
| LC: lymphoid cells | T-LBL: T-lymphoblastic lymphoma/leukemia |

## II. Age-specific incidences of acute lymphoblastic leukemia (ALL) and chronic lymphocytic leukemia (CLL)

ALL and CLL are two forms of lymphoid leukemia, but they have big difference on pathology, occurring age, and prognosis. ALL occurs mostly in children and proceeds quickly, whereas CLL occurs mainly in old people and proceeds slowly. However, two sub-types of ALL, namely the Ph$^+$-ALL and the Ph-like ALL, occur in both children and adults (Stephen, 2015).

**2.1 ALL has a peak incidence in young children**

ALL occurs at any age, but mostly in children. 80% of cases of childhood leukemia are ALL. ALL has a peak incidence at age 2-5 (Figure 1). The incidence of ALL is low after age 25. Boys have higher incidence of ALL than girls by a factor of 4:3 (Hutter, 2010). ALL arises from a lymphoblast or a pro-lymphocyte. In pathology, ALL has three subtypes: L1, L2, and L3. These subtypes of ALL are distinguished by size of leukemia cells: small cells in ALL-L1, large cells in ALL-L3, and mixture of small and large cells in ALL-L2. Among these three subtypes, ALL-L1 has the best prognosis (MoradiAmin, 2016).

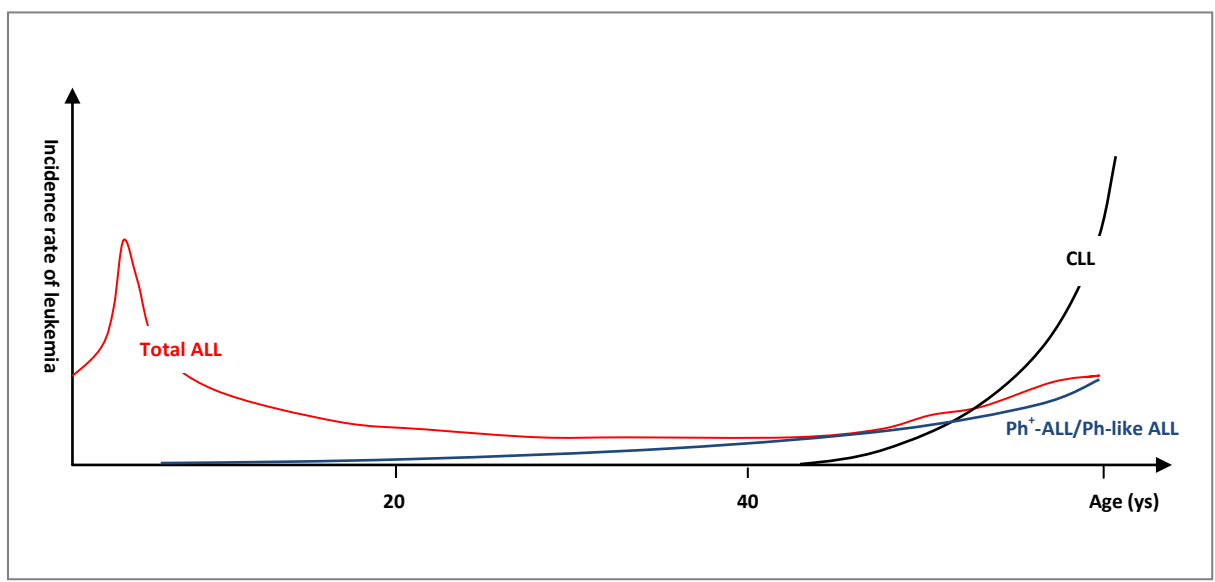

Figure 1. Age-specific incidences of CLL and subtypes of ALL (a schematic graph)

ALL and CLL are both lymphoid leukemia, but they are quite different on pathology and occurring age. ALL occurs at any age but has a peak incidence at age 2-5 (**red line**). The total incidence of ALL is low after age 25. Ph$^+$-ALL and Ph-like ALL are two subtypes of ALL, and they occur in older children and adults (**blue line**). CLL affects mainly old people (**black line**).

Chromosome changes (CCs) are found in leukemia cells in 70% of pediatric ALLs (< age 15). Frequent forms of CCs in pediatric cases include aneuploid, t(12;21), and t(9;22) (Ph translocation). 50% of pediatric ALLs have aneuploid, 25% have t(12;21), and 3%-5% have Ph translocation (Table 1). In adult ALLs, 39% have complex karyotypes and 25% have Ph translocation. Since the ALLs that have Ph translocation have a distinct disease-presentation from other ALLs, they are named as Ph$^+$-ALL (Bleckmann, 2016). Some Ph-negative ALLs develop in a similar manner to Ph$^+$-ALL, and they are named as Ph-like ALLs (Roberts, 2014). The cases of Ph$^+$-ALL and Ph-like ALL account for 15% of pediatric ALLs and over 50% of adult ALLs. In a study with 148 adult ALLs, 33% of cases are Ph-like ALL, 31% are Ph$^+$-ALL, and 36% are B-other ALL (Jain, 2017). Table 1 shows the differences between subtypes of ALL on occurring age, driver DNA changes, and 5-year survival rate. B-other ALL occurs mainly in young children and has a peak incidence at age 2-5. The most frequent DNA changes in B-other ALL are aneuploid and t(12;21). The 5-year survival rate of B-other ALL after treatment is 95% in pediatric cases (age < 15) and 50% in adult cases (age > 15).

Table 1. Differences between subtypes of ALL on driver DNA changes and 5-year survival rate

|  | B-other ALL | Ph$^+$-ALL | Ph-like ALL | CLL |
|---|---|---|---|---|
| **Occurring age** | Peak at age 2-5 | > age 10 | > age 10 | > age 55 |
| **Increasing incidence with age** | No | Yes | Yes | Yes |
| **Driver DNA changes** | Aneuploid or t(12;21) | Ph translocation + gene mutations | CRLF2 rearrangement + gene mutations | Gene mutations |
| **5-year survival rate in age <15** | 95% | 85% | 92% | -- |
| **5-year survival rate in age >15** | 50% | 25% | < 20% | 50% |

## 2.2 Ph$^+$-ALL and Ph-like ALL both occur in older children and adults

Ph$^+$-ALLs account for 25% of adult ALLs but only 3% of pediatric cases (Bleckmann, 2016). Ph$^+$-ALL occurs rarely in children younger than age 10, but it has increasing incidence with age (Figure 1 and Table 1). Similar to that in chronic myeloid leukemia (CML), Ph translocation is responsible for Ph$^+$-ALL development. Ph translocation contributes to cell transformation of a lymphoblast by generating a *BCL-ABL1* fusion gene and permanently activating the tyrosine kinase of ABL1 (Bennour, 2016). Notably, in Ph-translocation-related cell transformation, other DNA changes are also needed. For example, in some Ph$^+$-ALLs, activation of SRC kinase is also required for cell transformation of lymphoblast. Importantly, Ph$^+$-ALL has poorer prognosis than B-other ALL (Yanada, 2015). The prognosis of Ph$^+$-ALL is worse in adult cases than in pediatric cases. The 5-year survival rate of Ph$^+$-ALL is 85% in pediatric patients but only 25% in adult patients (Table 1) (Chiaretti, 2015). Adult Ph$^+$-ALLs have high rate of relapse after treatment. Since activation of the tyrosine kinase of ABL1 by Ph translocation is associated with the poor prognosis of Ph$^+$-ALL, tyrosine kinase inhibitors (TKI) can be applied in Ph$^+$-ALL treatment. Clinic data have shown that TKIs including Imatinib are effective agents in controlling the proliferation of leukemia cells in CML (Saini, 2017).

Ph-like ALL is another subtype of ALL, which has similar development features to Ph$^+$-ALL. Ph-like ALL occurs in older children and adults, and the incidence increases with age. Ph-like

ALLs account for 10% of pediatric ALLs and 27% of adult cases. Ph-like ALL cells do not have Ph chromosome but have other forms of DNA changes (Roberts, 2014). The most frequent forms of DNA changes in Ph-like ALL are CRLF2 rearrangements (Jain, 2017). Two major forms of CRLF2 rearrangements are IGH/CRLF2 translocation (in 76% of Ph-like ALLs) and P2RY8/CRLF2 translocation (17%). CRLF2 rearrangement results in over-expression of *CRLF2* gene and abnormal activation of JAK kinase. Ph-like ALLs have often poor prognosis, thus CRLF2 rearrangements can be used as an independent prognostic factor (Herold, 2017). Adult Ph-like ALLs have poorer prognosis than pediatric cases (Tran, 2016). The 5-year survival rate of Ph-like ALL after treatment is 92% in children and 20% in adults (Table 1).

## 2.4 CLL occurs mainly in old people

CLL affects mainly old people, and the incidence of CLL increases quickly after age 50 (Figure 1). Males have about two times of incidence of CLL as females (Passegue, 2002). 95% of CLLs are B-cell CLL. CLL arises from a mature B-cell, such as naïve B-lymphocyte and memory B-cell. CLL and small lymphocyte lymphoma (SLL) are now classified into the same entity, because they may have the same cell-of-origin and similar pathogenesis. Gene mutations are the main forms of DNA changes in CLL. However, all the mutations have low recurrences (Amin, 2016). The mutations with relative higher frequencies in CLL include mutations of *TP53, BIRC3, NPTCH1,* and *SF3B. TP53* and *BIRC3* mutations are associated poor prognosis of CLL. CCs are also observed in CLL, and they include trimosy 12, del (13q), del (11q), and del (17p). However, CCs are more often seen in late-stage CLLs (Puiggros, 2014; Karnolsky, 2000). CLL is a slow progressing disease, and the 5-year survival rate is over 50% (Table 1).

## III. Three potential sources of cell injuries of lymphoid cells (LCs)

To understand childhood leukemia and lymphoma, we have recently proposed two hypotheses on the causes and the mechanism of transformation of a LC (Wang-Michelitsch, 2018a). These hypotheses are: **A.** repeated bone-remodeling during bone-growth and bone-repair may be a source of cell injuries of hematopoietic cells in marrow; and **B.** a LC may have three pathways on transformation: a slow, a rapid, and an accelerated. Before discussing ALL and CLL, we firstly make a brief introduction of these hypotheses. In this part, we introduce the potential sources of cell injuries of LCs. In next two parts, we will introduce the hypothesized three pathways of cell transformation of a LC.

Cell injuries and DNA injuries are triggers for generation of DNA changes in cells. For LCs, there may be three sources of cell injuries: repeated bone-remodeling during bone-growth and bone-repair, for the LCs in marrow; long-term thymic involution, for the developing T-cells in thymus; and pathogen-infections, for the LCs in lymph nodes (LNs) and lymphoid tissues (LTs).

## 3.1 Repeated bone-remodeling during bone-growth and bone-repair in marrow cavity

In humans, hematopoiesis takes place mainly in marrow cavities of long bones and spongy parts of all bones. Bone-growth is a result of repeated modeling-remodeling of bone tissues. The marrow cavity and spongy part of a bone are developed and enlarged with the growth of the bone. Bone-remodeling is a process of absorption of the bone-tissues exposed to marrow cavity by osteoclasts. Osteoclasts digest bone tissues by secreting acid substances and enzymes. Thus, bone-remodeling may produce damaging substances to marrow cells including hematopoietic stem cells (HSCs), LCs, and myeloid cells (MCs). Namely, these hematopoietic cells have a risk to be injured by bone-remodeling. Although this risk is low, repetition of bone-remodeling during bone-development in a child can largely increase this risk. Additional bone injuries may disturb bone-growth and increase this risk. The children at age 2-5 have the highest incidence of bone fractures due to their weak muscles and bones. Thus, bone injuries in young children caused by frequent body movements may be related to the peak incidence of ALL at age 2-5. Since T-cells develop in thymus, bone-remodeling affects mainly developing B-cells, developing MCs, and HSCs

### 3.2 Long-term thymic involution in thymus

Thymus is an organ for T-cell development. This small organ is located in the mediastinum between two lungs. However, thymus undergoes involution since age of puberty and continues for many years. Thymic shrinking is a result of death of stromal cells. Death of a large number of stromal cells may produce toxic substances to the developing T-cells in thymus. Although this risk is low, constant thymic involution for many years can increase this risk. T-lymphoblastic lymphoma/leukemia (T-LBL) is a form of aggressive lymphoma that occurs mainly in adolescents and young adults (You, 2015). Mediastinal mass is the first syndrome in 60%-80% of T-LBLs. Thus, it is quite possible that T-LBL originates from a T-lymphoblast in thymus. Thymic involution may be related to the cell transformation of T-lymphoblast in T-LBL development.

Taken together, repeated bone-remodeling and constant thymic involution may be two internal damaging factors for LCs. Cell injuries by internal damage can occur also to other cells including tissues cells during body development and during inflammations. DNA changes can be generated also in injured tissue cells. However, cell transformation of a tissue cell is a slow process, as a co-effect of external and internal damaging factors. Differently, cell transformation of a LC can occur at young age as that seen in ALL. Thus, the effect of internal damage on LCs can be recognized.

### 3.3 Pathogen-infections in lymph nodes (LNs) and lymphoid tissues (LTs)

For the LCs that have passed LNs and/or LTs, pathogen-infections should be the main cause for cell injuries and DNA injuries. Bacterial and viral can both damage host cells, but viral may be stronger than bacterial on triggering cell transformation. Viral can proliferate inside host cells and cut host DNAs directly. Three major types of pathogens associated with lymphoma development are Epstein-Barr virus (EBV), human T-cell lymphotropic virus type 1 (HTLV-1), and *Helicobacter-Pylori* (*H. pylori*) (Geng, 2015, Oliveira, 2017, and Krishnan, 2014). Notably, not all the individuals that have had infections of these pathogens develop lymphoma. Repetition of infections seems to be more critical for causing cell transformation of a LC. Except in big epidemic, repetition of infections is often a consequence of immunodeficiency. Therefore, immunodeficiency increases the risk of lymphoma development by permitting the repetition of infections.

## IV. DNA changes are generated and accumulate in cells as a result of repeated cell injuries and repeated cell proliferation

DNA changes are generated in cells as a consequence of cell injuries and DNA injuries. There are two major types of DNA changes: point DNA mutation (PDM) and chromosome change (CC). CCs include numerical CCs (NCCs) and structural CCs (SCCs). A NCC exhibits as gain or loss of one or more chromosomes of a cell. A SCC exhibits as translocation (t), deletion (del), gain (+), or inversion (inv) of part of a chromosome. Different types of DNA changes are generated by different mechanisms. Studies show that generation of PDM is a result of Misrepair of DNA on a double-strand DNA break (Natarajan, 1993; Bishay, 2001; Kasparek, 2011). Namely, a PDM is not "DNA damage", but rather is a result of incorrect repair of "DNA damage" (Wang-Michelitsch, 2015).

Similarly, a SCC is also generated as a result of Misrepair of DNA. A SCC can be generated when a cell has multiple DNA breaks. For example, chromosome translocation is a result of re-linking of a broken DNA by a "wrong" DNA fragment, which is released from another chromosome by DNA breaks. Differently, generation of a NCC is often a consequence of dysfunction of cell division. A PDM/SCC is generated for DNA repair, and it is essential for maintaining the structural integrity of DNA and for preventing cell death. Hence, generation of PDM/SCC is not really a mistake. However, a NCC is not generated for "repair". Thus survival of a cell with a NCC is a real mistake. Unfortunately, a NCC can affect multiple genes and may cause cell transformation. Some pediatric ALLs may develop by such a mistake, because the driver DNA change in these cases is aneuploid, which is a form of NCC (Wang-Michelitsch, 2018a).

Accumulation of DNA changes in cells is a result of repeated cell injuries and repeated cell proliferation. For example, in marrow cavity, regeneration of HSCs and proliferation of developing LCs enables the accumulation of DNA changes in HSCs and in LCs. In LNs/LTs, chronic infections drive the accumulation of DNA changes in LCs by causing cell injuries and promoting cell proliferation of LCs. HSCs and memory cells are stem cells that are regenerable for our whole lifetime. Thus, only the DNA changes that are generated or inherited in HSCs and memory cells can accumulate for a long time. By permitting repeated infections, immunodeficiency may accelerate the accumulation of DNA changes in LCs.

## V. A LC may have three pathways on cell transformation: a slow, a rapid, and an accelerated

CCs can be classified into three groups by their effects on a LC: great-effect CCs (GECCs), mild-effect CCs (MECCs), and intermediate-effect CCs (IECCs) (Wang-Michelitsch, 2018b).

A PDM and a MECC are often mild to cells, thus they can accumulate in cells. Some PDMs/MECCs may contribute to cell transformation. Differently, a GECC affects one or more genes and can alone drive cell transformation. An IECC affects one or more genes and contribute to cell transformation. The IECC-associated cell transformation is a co-effect of IECC(s), and PDMs/MECCs. With stronger effect than a PDM/MECC, an IECC can accelerate the cell transformation driven by accumulation of PDMs and MECCs.

A LC is different from a tissue cell on some aspects. A LC may have higher survivability from DNA changes than a tissue cell by three cellular characteristics: anchor-independence on survival, inducible expression of cell surface molecules, and expression of fewer genes as being the smallest cell. However, among all LCs, differentiating LCs may have higher tolerance to DNA changes than non-differentiating LCs. Differentiating LCs include progenitor cells, blast cells, and pro-cytes. Non-differentiating LCs include naïve lymphocytes, memory cells, and effector cells. The higher tolerance to DNA changes makes a LC have a risk to be transformed by CCs. In addition, a LC has by nature some properties similar to that of a cancer cell. These properties include anoikis-resistance, non-inhibition by cell-contact, production of matrix metalloproteinases, and mobility (Wang-Michelitsch, 2018b). Thus, for cell transformation, a LC requires obtaining fewer cancerous properties than a tissue cell. Namely, a LC needs to acquire only one property for transformation: the stimulator-independent mitosis. This means that a LC can be transformed more rapidly, even in "one step". Therefore, a LC can be transformed not only by accumulation of PDMs, MECCs, and IECCs, but also possibly by a GECC in "one step".

Based on the above analysis, we hypothesized that a LC may have three pathways on transformation (Wang-Michelitsch, 2018b). These pathways are: **A.** a slow one driven by accumulation of PDMs and MECCs through many generations of cells; **B.** a rapid one driven by a GECC in "one step" in one generation of cell; and **C.** an accelerated one driven by accumulation of PDMs, MECCs, and IECC(s) through a few generations of cells (Box 1). Notably, a tissue cell and a non-differentiating LC cannot survive from a GECC and an IECC, thus they can only be transformed via slow pathway. Differently, a differentiating LC can be transformed via all three pathways. In slow and accelerated pathways, long-term accumulation of PDMs, MECCs, and/or IECCs takes place mainly in long-living stem cells including HSCs and memory cells. Only the final PDM/MECC/IECC is generated in the first transformed cell. Cell transformations via different pathways occur at different ages. A cell transformation via slow pathway takes places mainly in adults and has an increasing incidence with age. A cell transformation via rapid pathway can take place at any age and has no increasing incidence with age. A cell transformation via accelerated pathway can take place also at any age, but it has increasing incidence with age (Wang-Michelitsch, 2018b).

Box 1. Hypothesized three pathways of a LC on cell transformation

- **Slow pathway**: by accumulation of PDMs and MECCs through many generations of cells
- **Rapid pathway**: by a GECC in "one step" in one generation of cell
- **Accelerated pathway**: by accumulation of PDMs, MECCs, and IECC(s) through a few generation of cells

PDM: point DNA mutation  GECC: great-effect chromosome change
MECC: mild-effect chromosome change  IECC: intermediate-effect chromosome change

## VI. The age of occurrence of leukemia is determined by the transforming pathway of a LC

Distinguishing between three pathways of cell transformation of a LC can help us to understand why some forms of lymphoid leukemia and lymphoma occur mainly in children and why they are mostly aggressive. In this part, we discuss how the age of occurrence of leukemia is determined by the transforming pathway of LC.

### 6.1 A leukemia occurring mainly in adults: via slow pathway

Most forms of cancers occur mainly in adults, and their incidences increase with age. The reason is: the cell transformations in these cancers are results of long-term accumulation of PDMs and MECCs through many generations of cells (Wang-Michelitsch, 2015). A leukemia occurring mainly in adults is also a result of cell transformation of a LC by accumulation of PDMs and MECCs via slow pathway. All LCs can be transformed via slow pathway. A naïve lymphocyte and a memory cell can be transformed mainly via this pathway. When a lymphoblast is transformed via slow pathway, lymphoid proliferative disorder (LPD) may occur. When a naïve or memory cell is transformed, CLL may occur (Figure 2). A tissue cell has low tolerance to DNA changes, thus it can be transformed only by accumulation of PDMs and MECCs. This means that solid cancers and sarcomas develop mostly via slow pathway. Occurring only in adults, CLL develops mainly via this pathway. Having a peak incidence at age 2-5, B-other ALL may not develop via this pathway. $Ph^+$-ALL and Ph-like ALL occur at all ages, thus they may not develop via this pathway.

### 6.2 A leukemia occurring at any age without increasing incidence with age: via rapid pathway

Having higher survivability from DNA changes and requiring obtaining fewer cancerous properties for transformation than a tissue cell, a lymphoblast/pro-lymphocyte has a risk to be transformed by a GECC in only "one step", namely in one generation of cell. The GECC is generated in the first transformed cell. A cell transformation via this pathway can take place at any age (Figure 2). Such a rapid transformation does not require accumulation of DNA changes; thus the incidence does not increase with age. Since only a differentiating LC can be transformed via this pathway and a GECC may affect cell differentiation, rapid transformation often leads to occurrence of acute leukemia such as ALL. Having a peak incidence in young children, B-other ALL develops more likely via this pathway. $Ph^+$-ALL and Ph-like ALL may not develop via this pathway, because they have increasing incidences with age. Not occurring in children, CLL may not develop via this pathway. The rapid pathway of transformation of a lymphoblast makes us understand why leukemia can occur in young children and why pediatric leukemias are mostly acute.

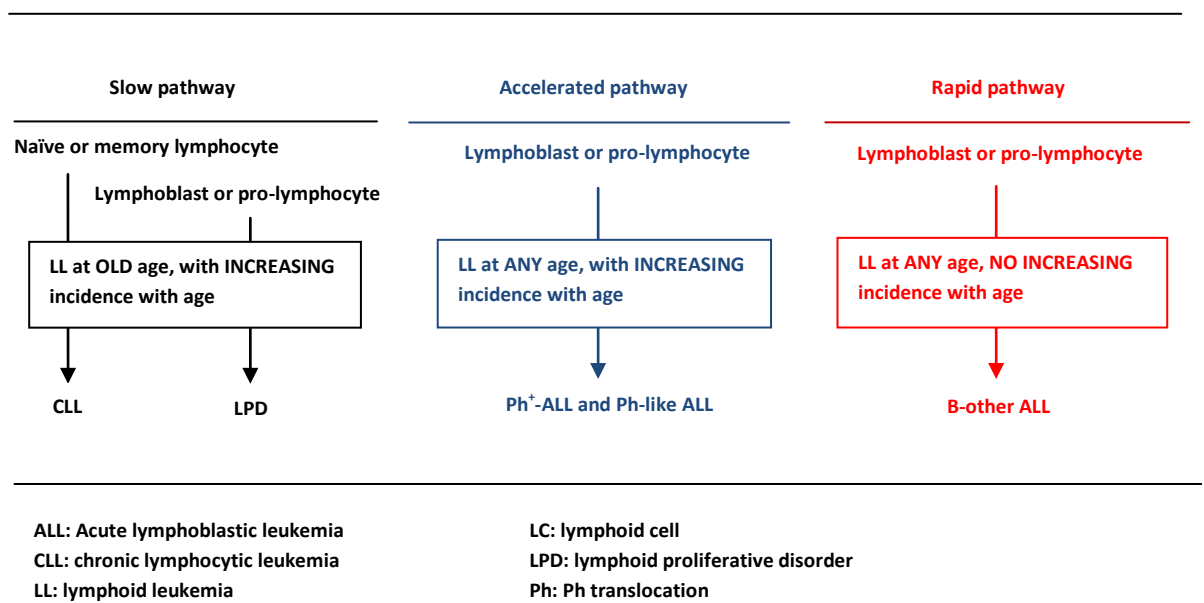

Figure 2. The age of occurrence of lymphoid leukemia is determined by the transforming pathway of LC

A LC may have three pathways on cell transformation: a **slow**, a **rapid**, and a**n accelerated**. Cell transformations via different pathways occur at different ages. A transformation via slow pathway takes places mainly in adults and has increasing incidence with age. A transformation via rapid pathway can take place at any age and has no increasing incidence with age. A transformation via accelerated pathway can take place also at any age, but it has increasing incidence with age. Thus, CLL may develop as a result of transformation of a naïve lymphocyte or memory cell via slow pathway. Lymphoid proliferative disorder (LPD) may occur when a lymphoblast is transformed via slow pathway. B-other ALL may develop as a result of transformation of a lymphoblast/pro-lymphocyte via rapid pathway. $Ph^+$-ALL and Ph-like ALL may both develop as a result of transformation of a lymphoblast/pro-lymphocyte via accelerated pathway.

## 6.3 A leukemia occurring at any age with increasing incidence with age: via accelerated pathway

$Ph^+$-ALL and Ph-like ALL occur at all ages and their incidences increase with age. So they develop more likely via accelerated pathway (Figure 2). The cell transformation via this pathway is driven by IECC(s) and PDMs/MECCs through some generations of cells. Ph translocation and CRLF2 rearrangements are the IECCs respectively in Ph$^+$-ALL and Ph-like ALL. Since an IECC can accelerate cell transformation, the transformation via accelerated pathway can occur at young ages including children and young adults. In addition, an IECC may have increasing occurrence in LCs with age of our body. Thus, the IECC-driven cell transformation has increasing incidence with age. A lymphoblast/pro-lymphocyte may be tolerant to an IECC and transformable via this pathway. Cell transformation of a lymphoblast via accelerated pathway can lead to occurrence of acute leukemia. B-other ALL has a peak incidence in young children, thus it may not develop via this pathway. CLL does not occur in young ages, thus it may not develop via this pathway.

## VII. Development of pediatric ALL may be related to the repeated bone-remodeling during bone-growth

In the above discussion, we have briefly introduced our hypotheses on the causes of cell injuries of LCs and on the pathways of cell transformation of a LC. Our hypotheses suggest that the age of occurrence of lymphoid leukemia is mainly determined by the transforming pathway of a LC. In this part, we analyze the developing characteristics of ALL and CLL by these hypotheses.

### 7.1 Childhood ALL: as a result of rapid cell transformation of a lymphoblast/pro-lymphocyte

ALL occurs mostly in children. Over 70% of ALLs occur before age 20. Only 3% of pediatric ALLs (< age 15) are Ph$^+$-ALL and 10% are Ph-like ALL. Thus, most ALLs occurred before age 10 are B-other ALL, which develops by rapid cell transformation of a B-lymphoblast/pro-lymphocyte. Aneuploid and t(12;21) are two major forms of CCs in pediatric ALLs, thus they may be the GECCs that are responsible for the rapid transformation of a lymphoblast in B-other ALL (Table 2). In marrow, lymphoblast and pro-lymphocytes are in large numbers and are in processes of cell division. These cells have high risk to be injured during cell division, thus they have a risk of generation of NCCs. The high frequency of aneuploid in ALLs is a proof. By causing cell injuries of developing B-LCs in marrow, repeated bone-remodeling during bone-growth may be a trigger for B-ALL development in a child. T-cells develop in thymus, thus bone-remodeling does not affect developing T-cells. This explains why 85% of childhood ALLs are B-cell ALL.

ALL has the highest incidence at age 2-5. In our view, age 2-5 may be a peak age of bone-remodeling by two factors: **A.** age 0-10 is the peak age of formation of cavities of small bones; and **B.** age 2-5 is the peak age of occurrence of bone-fractures in children. It is known that different bones start to develop at different ages. Most small long bones start to ossification only after birth. Till age 10, all small bones will have a marrow cavity (Horton, 1990). Thus, age 0-10 may be a peak time for bone-remodeling of small bones. In addition, young children at age 2-5 have the highest occurrence of bone-fractures, because they have weak muscles and bones. Children may have sometime asymptotic micro-injuries in bones. Bone injuries may disturb bone-growth and increase the risk of injuries of hematopoietic cells by bone-remodeling. Young boys may have higher risk of bone-injuries than girls by more frequent body movements. This may explain the higher occurrence of ALL in boys than girls. Taken together, the high incidence of ALL in young children may be a co-effect of two factors: the probability of rapid transformation of a lymphoblast and the higher risk of cell injuries of developing LCs by repeated bone-remodeling in children than in adults.

**Table 2. Types of DNA changes in CLL and subtypes of ALL**

|  | Driver DNA changes | Cell of origin | Potential source of cell injuries | Pathway of cell transformation | Age of occurrence |
|---|---|---|---|---|---|
| B-other ALL | Aneuploid or t(12;21) | Lymphoblast or pro-lymphocyte | Repeated bone-remodeling during bone-growth | Rapid | Any age |
| $Ph^+$- ALL | Accumulation of PDMs/MECCs +Ph translocation | Lymphoblast or pro-lymphocyte | Repeated bone-remodeling in marrow | Accelerated | Any age |
| Ph-like ALL | Accumulation of PDMs/MECCs /IECC(s) | Lymphoblast or pro-lymphocyte | Repeated bone-remodeling in marrow | Accelerated | Any age |
| CLL | Accumulation of PDMs/MECCs | Naïve B-cell or Memory cell | Repeated bone-remodeling in marrow + pathogen-attacking in lymph tissues | Slow | > age 50 |

In pathology, ALL has three subtypes: L1, L2, and L3. These subtypes have different prognoses: ALL-L1, the subtype with small leukemic cells, has the best prognosis; whereas ALL-L3, the subtype with large cells, has the worst prognosis. ALL-L1 and ALL-L3 have different prognoses, probably because they have different cells-of-origin and different grades of cell transformation. During the development of lymphocytes, LCs at different stages have different sizes: a lymphoblast is bigger than a pro-lymphocyte, and a pro-lymphocyte is bigger than a naïve lymphocyte. ALL-L1 arises more likely from a pro-lymphocyte whereas ALL-L3 arises probably from a lymphoblast. A lymphoblast has lower maturity on cell functions than a pro-lymphocyte, thus transformation of a lymphoblast will lead to occurrence of a more aggressive subtype of ALL (such as ALL-L3). In addition, ALL-L3 may develop as a result of high-grade cell transformation of a lymphoblast, but ALL-L1 may result from intermediate-grade cell transformation.

## 7.2 Ph+-ALL and Ph-like ALL: as consequences of cell injuries of hematopoietic stem cell (HSCs) and LCs in marrow

Occurring in children and adults, $Ph^+$-ALL and Ph-like ALL develop more likely via accelerated pathway. A cell transformation via accelerated pathway is a co-effect of IECC(s) and accumulation of PDMs/MECCs (Table 2). Ph translocation is the IECC in $Ph^+$-ALL, and CRLF2 rearrangement is the IECC in Ph-like ALL. Since accumulation of PDMs/MECCs occurs mainly in precursor HSCs, cell injuries of HSCs are also associated with the development of Ph$^+$-ALL and Ph-like ALL. Notably, Ph$^+$-ALL and Ph-like ALL have worse prognoses than B-other ALL. The difference on prognosis between Ph$^+$-ALL/Ph-like ALL and B-other ALL may be related to their difference on the transforming pathway of lymphoblast: the former is via accelerated pathway and the latter is via rapid pathway.

The cell transformation of lymphoblast in Ph$^+$-ALL or Ph-like ALL is driven by accumulation of PDMs, MECCs, and IECC(s) through some generations of cells. Thus, in a patient with Ph$^+$-ALL or Ph-like ALL, the "sister cell" and some of "cousin cells" of the first transformed cell are a kind of pre-cancer cells. The reason is that these cells have similar DNA changes to this transformed cell, because they all have the same precursor HSCs and LCs. Chemotherapy and radiotherapy may promote cell transformation of one of the pre-cancer cells (Rothkamm 2002). Thus, existence of pre-cancer cells may be a reason for the cancer relapse in Ph$^+$-ALLs and Ph-like ALLs. In addition, adult cases of Ph$^+$-ALL and Ph-like ALL have worse prognosis than pediatric cases. This may be also related to the pre-cancer cells. The pre-cancer cells in adult cases of Ph$^+$-ALL and Ph-like ALL should have more PDMs/MECCs than pediatric cases, thus they have higher risk to be transformed by chemotherapy. Differently, in B-other ALL, the rapid cell transformation takes place in "one step" by a GECC. The GECC is generated in the first transformed cell. Thus, in B-other ALL, there exists no pre-cancer cell.

### 7.3 CLL: as a consequence of cell injuries of LCs in LNs/LTs and that of HSCs/LCs in marrow

CLL develops as a result of cell transformation of a LC driven by accumulation of PDMs and MECCs (Table 2). CLL is an indolent form of leukemia, indicating that it originates from a naïve or memory B-cell. However, CLL may arise more often from a memory B-cell. A memory cell is a downstream cell of a naïve lymphocyte, thus the former has always more number of PDMs/MECCs than the latter. Since accumulation of PDMs and MECCs proceeds mainly in HSCs and memory cells, the cell injuries of LCs (including memory cells) occurred in LNs/LTs and that of HSCs/LCs occurred in marrow cavity may be all related to cell transformation of a memory B-cell. Therefore, CLL develops as a co-effect of multiple factors. Probably, repeated pathogen-infections, repeated bone-remodeling during bone-growth and bone-repair, and repeated exposures to radiation/chemicals may all contribute to CLL development. Males have much higher incidence of CLL than females. A reason may be: men have often higher risk of bone injuries than women, because they have heavier body weights and heavier physical works including sports.

### VIII. Conclusions

We have discussed in this paper the causing factors and the mechanism of cell transformation of LCs in developments of ALL and CLL. We show that the ALL in young children may be a result of "one-step" transformation of a lymphoblast/pro-lymphocyte. Ph$^+$-ALL and Ph-like ALL may be a result of cell transformation of a lymphoblast via accelerated pathway. CLL is often a result of transformation of a memory B-cell via slow pathway. By causing cell injuries of HSCs and LCs in marrow, repeated bone-remodeling during bone-growth may be a trigger for the cell transformation of a lymphoblast in a child. Differently, CLL develops by multiple factors, probably including pathogen-infections, exposures to radiation/chemicals, and/or repeated bone-remodeling during bone-growth and bone-repair. To verify our hypothesis that repeated bone-remodeling is related to ALL development, experimental researches can be undertaken to study the association of repeated bone injuries with the development of leukemia in animal models (such as rabbits).

Occurrence of ALL in a child is a catastrophe. This catastrophe may be a co-effect of three factors: **A.** risk of cell injuries of hematopoietic cells by repeated bone-remodeling during bone-growth before age 25; **B.** higher tolerance of a developing LC from DNA changes than a tissue cell; and **C.** requiring obtaining fewer cancerous properties of a developing LC for cell transformation than a tissue cell. Hence, development of ALL can be understood as a "side-effect" of bone growth and functionality of lymphocytes. These factors are unfortunately internal and unavoidable. To reduce the risk of ALL development, avoiding violent sports and bone-injuries may be helpful. Cryopreservation of HSCs in umbilical cord at birth is also advised.